\documentstyle[sprocl]{article}
\input epsf

\input{psfig}

\def\Journal#1#2#3#4{{#1} {\bf #2}, #3 (#4)}

\def\NPB{{\em Nucl. Phys.} B}
\def\NPA{{\em Nucl. Phys.} A}
\def\PLB{{\em Phys. Lett.}  B}
\def\PRL{\em Phys. Rev. Lett.}
\def\PRD{{\em Phys. Rev.} D}
\def\PRC{{\em Phys. Rev.} C}

\def\half{{1\over 2}}
\def\MeV{{\rm MeV}}

\def\Tr{{\rm Tr\,}}
\def\nrcpt{NR\raise.4ex\hbox{$\chi$}PT\ }

\def\ltap{\ \raise.3ex\hbox{$<$\kern-.75em\lower1ex\hbox{$\sim$}}\ }
\def\gtap{\ \raise.3ex\hbox{$>$\kern-.75em\lower1ex\hbox{$\sim$}}\ }

\def\CA{{\cal A}}

\def\CL{{\cal L}}

\def\pds{{\it PDS}\ }

\def\bfq{{\bf q}}

\def\frac#1#2{{\textstyle{#1\over#2}}}
\def\darr#1{\raise1.5ex\hbox{$\leftrightarrow$}\mkern-16.5mu #1}
\def\){\right)}
\def\({\left(}
\def\]{\right]}
\def\[{\left[}
\def\si{{}^1\kern-.14em S_0}
\def\siii{{}^3\kern-.14em S_1}
\def\diii{{}^3\kern-.14em D_1}
\def\fm{{\rm\ fm}}
\def\MeV{{\rm\ MeV}}
\def\CA{{\cal A}}
\def\Czzm{ {\cal A}_{-1[00]} }
\def\Cttm{{\cal A}_{-1[22]} }
\def\Ctzm{{\cal A}_{-1[20]} }
\def\Cztm{ {\cal A}_{-1[02]} }
\def\Czzz{{\cal A}_{0[00]} }
\def\Cttz{ {\cal A}_{0[22]} }
\def\Ctzz{{\cal A}_{0[20]} }
\def\Cztz{{\cal A}_{0[02]} }

\def\be{\begin{equation}}
\def\ee{\end{equation}}
\def\bea{\begin{eqnarray}}
\def\eea{\end{eqnarray}}

\begin{document}

\title{INCLUDING PIONS\footnote{Talk presented at the Joint Caltech/INT workshop
on {\bf Nuclear Physics with Effective Field Theory}, Caltech, February 1998.}
\footnote{NT@UW-11}
}

\author{{Martin  J. Savage
\footnote{Work done in collaboration with David B. Kaplan and Mark B. Wise }
}}

\address{
Department of Physics, University of Washington,  Seattle, WA 98195}

\maketitle
\abstracts{
Recent progress in using effective field theory 
to describe systems  with two nucleons is discussed 
with particular emphasis placed on the inclusion
of pions.
Inconsistencies arising in Weinberg's power counting 
are demonstrated with two concrete examples.
A consistent power-counting scheme is discussed in which pion exchange 
is sub-leading to local four-nucleon operators.
NN scattering in the $\si$ and $\siii-\diii$ 
channels is calculated at sub-leading order 
and compared with data.
}

\section{Introduction}
\label{sec:1}

The last several years have seen significant progress towards
an effective field theory description of  the interactions 
between nucleons.
The ultimate goal of this endevour is to construct a framework
with which to describe
multi-nucleon systems, both bound and unbound, and 
inelastic processes in a systematic way.
This effort was initiated by Weinberg's pioneering work on the 
subject \cite{Weinberg1}\ where he proposed a power-counting
scheme for local-operators involving two or more nucleons
and the inclusion of pions.
This proposal had many important implications from detailing 
how to include higher order terms in the chiral and derivative expansion
to the prediction of small three-body forces.
Applications to phenomenology, such as studies of the $NN$ phase-shifts,
$np\rightarrow D\gamma$ radiative capture,  nucleon-deuteron scattering,
as well as extensive discussion of the underlying field theory itself
have followed Weinberg's original proposal\cite{KoMany}$^-$\cite{steelea}.
Problems with Weinberg's power-counting  became apparent when 
computations were performed with dimensional regularization\cite{KSWa}, 
as opposed to a  momentum-space regulator that previous 
computations had employed\cite{KoMany}.
Contributions to $NN$ scattering in the $\siii-\diii$ channel
that are leading order in  Weinberg's power-counting
require counterterms at all orders in the momentum expansion, and 
therefore terms that are naively subleading order
must be included at leading order.
One is lead to the conclusion that Weinberg's power-counting is not consistent.

Recently, a new and consistent 
power-counting scheme has been proposed\cite{KSWb}\ 
that alleviates the difficulties encountered with Weinberg's 
power-counting.
It is this power-counting and its implications that I will present  
with an emphasis on  the theory with  pions.
Firstly, I will discuss Weinberg's power-counting scheme 
and its shortcomings.
Next, I will discuss a new power-counting scheme\cite{KSWb}
and show results from its application
to $NN$ scattering in the $\si$ and $\siii-\diii$ channels at subleading order.

\section{Weinberg's Power Counting}
\label{sec:2}

Effective field theories (EFT) are constructed to reproduce all S-matrix elements
with external momenta $\sim Q$ less than some scale $\Lambda$.
The scale $\Lambda$ is typically set by the mass of particles and kinematic regimes
not explicitly included in the theory as dynamical degrees of freedom.
However, the effect of these  particles and regimes  are included in the EFT by the 
presense of higher dimension operators, with coefficients explicitly set by 
$\Lambda$.
In addition to forming the lagrange density consistent with the global and local 
symmetries of the 
underlying theory, one must also define the method of regulating divergences.
It is desirable to choose a regulator that preserves the symmetries of the theory
and also maintains the hierarchy of the higher dimension operators.
Here we reproduce Weinberg's power-counting scheme\cite{Weinberg1,KSWa}, 
without an explicit discussion
of the method of regularization, therefore the conclusions are regulator independent.

The most general lagrange density consistent with chiral symmetry describing the
interaction of two nucleons is 
\bea
 \CL & = &  N^\dagger (iD_0+ \vec D^2/2M) N 
 + {f^2\over 8} \Tr \partial_\mu\Sigma^\dagger  \partial^\mu \Sigma +
{f^2\over 4}\omega\Tr m_q (\Sigma + \Sigma^\dagger)  
\nonumber\\
& -& \half C_S (N^\dagger N)^2 
-\half C_T (N^\dagger \vec\sigma N)^2 
+ g_A N^\dagger \vec A\cdot\vec\sigma N
\ \ +\ \ ...
\ \ \ ,
\label{eq:weinL}
\eea
where $D_\mu$ denotes a chiral-covariant derivative and
$\Sigma$ is the exponential of the isotriplet of pions
\bea
\Sigma & = & \exp \left({2i\over f} M\right)
\nonumber\\
M & = & \left( 
\matrix{\pi^0/\sqrt{2} & \pi^+\cr  \pi^- & -\pi^0/\sqrt{2} }
\right)
\ \ \ ,
\eea
with $f=132\ {\rm MeV}$ the pion decay constant and 
$\vec A$ is the axialvector meson field.
The ellipses denote
terms with more spatial derivatives and also more insertions of the 
light quark mass matrix, $m_q$.
The Georgi-Manohar\cite{GeorgiManohar} 
naive dimensional analysis arising from a consideration of 
loop contributions to observables suggests that 
$C_{S,T}\sim 1/f^2$, 
and Weinberg's power-counting will follow directly.

\begin{figure}[t]
\psfig{figure=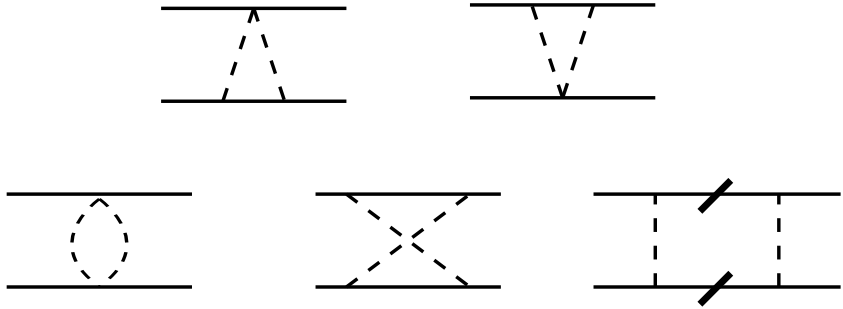,height=2.0in}
\caption{2-pion exchange Feynman graphs contributing to
the 2-nucleon potential $V^{(2)}$.  The first four are 2-nucleon irreducible;
the last diagram is 2-nucleon reducible, and the poles from the slashed
propagators are not included in the $\int {\rm d}q_0$ loop integration. 
\label{wein_fig1}}
\end{figure}

A necessary ingredient for an EFT is a power counting scheme that dictates
which graphs to compute in order to determine an observable 
to a desired order in the expansion.  
The main complication in the theory of nucleons and pions is
the fact that a nucleon propagator 
$S(q)=i/(q_0 - {\bf q}^2/2M)$ scales like $1/Q$ if $q_0$ scales like 
$m_\pi$ or an external 3-momentum,    
while $S(q)\sim M/Q^2$ if $q_0$ scales like an external kinetic
energy.  
Similarly, in loops $\int {\rm d} q_0$ can scale like $Q$ or $Q^2/M$,
depending on which pole is picked up. 
To distinguish between these two
scaling properties
it is convenient to  define generalized ``$n$-nucleon potentials'' $V^{(n)}$
comprised of those parts of connected Feynman diagrams with
$2n$ external nucleon lines that have no powers of $M$ in their scaling 
(except from relativistic corrections). 
Since there is no nucleon-antinucleon pair creation in the effective theory
such a  diagram always has exactly $n$ nucleon lines running through it.  
$V^{(n)}$ includes  diagrams which are $n$-nucleon irreducible 
and  parts of diagrams which are 1-nucleon irreducible.
To compute the latter contribution  to $V^{(n)}$
one identifies all combinations of two or more internal nucleon lines
that can be simultaneously on-shell, and excludes their pole
contributions when performing the $\int{\rm d}q_0$ loop integrations.  An
example of the 2-pion exchange contributions to $V^{(2)}$ is shown in 
Fig.(\ref{wein_fig1}).
A general $n$-nucleon Feynman diagram in the EFT  can be constructed by contracting 
the nucleon legs of $V^{(r)}$ potentials with $r\le n$.
Treating
the  $V^{(r)}$'s like vertices,
the $\int {\rm d}q_0$ loop integrations pick
up the poles of all the connecting nucleon lines.
The reason for this construction is that within the $V^{(r)}$ potentials,  all
nucleon propagators are off-shell and scale like $1/q_0\sim1/Q$. In contrast,
when one picks up the pole contribution from one of the  nucleon lines
connecting the $V^{(r)}$ ``vertices'', other nucleon lines will be almost
on-shell, and scale like $M/Q^2$.
A contribution to the $r$-nucleon
potential $V^{(r)}$ with $\ell$ loops, $I_n$ nucleon propagators, $I_\pi$ pion
propagators, and $V_i$ vertices involving $n_i$ nucleon lines and $d_i$
derivatives,  scales like $Q^{\mu}$, where
\begin{eqnarray}
\mu & = & 4 l - I_n - 2I_\pi + \sum V_i d_i
\ \ \ ,\nonumber\\
\ell & = & I_n + I_\pi - \sum V_i + 1
\ \ \ ,\nonumber\\
I_n+r & = & \half\sum V_i n_i 
\ \ \  .
\label{eq:weinrel}
\end{eqnarray}
In this power counting we take $m_\pi\sim Q$ and treat factors of the $u$ and
$d$ quark masses at the vertices as order $Q^2$. Combining these relations
leads to the scaling law for the $r$-nucleon
potential $V^{(r)}$ ($r\ge 2$):
\begin{eqnarray}\label{wpo}
\mu & = & 2+2\ell -r + \sum_i V_i (d_i+\half n_i-2)
\ \ \ .
\end{eqnarray}
Since chiral symmetry implies that the pion is derivatively coupled,  it
follows that $ (d_i+\half n_i-2)\ge 0$, which  implies that for a 2-nucleon
potential, $\mu\ge 0$, and that $\mu=0$ corresponds to tree diagrams.
It is straight forward to find the scaling property for a general Feynman
amplitude, by repeating the analysis that leads eq.(\ref{wpo})  treating the
$V^{(r)}$ potentials as $r$-nucleon vertices with $\mu$ derivatives, $\mu$
given by eq.(\ref{wpo}).  
While eq.(\ref{wpo})\ was derived assuming that $\int
{\rm d}q_0\sim Q$ and nucleon propagators scaled like $\sim 1/Q$, 
for these loop graphs they scale like $Q^2/M$ and $M/Q^2$ respectively.    
A general Feynman
diagram is constructed by stringing together $r$-nucleon potentials $V^{(r)}$.

Two nucleon scattering is simple since the
graphs are all ladder diagrams with insertions of  $V^{(2)}$'s acting
as ladder rungs.   
Each loop of the ladder introduces a loop integration (${\rm
d}q_0{\rm d}^3\bfq\sim Q^5/M$) and two nucleon propagators ($ M^2/Q^4$) to
give a  factor of $(QM)$ per loop.
Expanding
$V^{(2)} = \sum_{\mu=0}^\infty  V^{(2)}_\mu $,
where $V_\mu^{(2)}\sim Q^\mu$,
a 2-nucleon diagram whose $i^{th}$ rung is the generalized potential
$V^{(2)}_{\mu_i}$
scales as
\begin{eqnarray}\label{ourpo}
Q^\nu (QM)^L\ ,\qquad
\nu & = & \sum_{i=1}^{L-1} \mu_i
\ \ \ \ ,
\end{eqnarray}
where  $L$ is the number of loops external to
the $V^{(2)}$'s.
Since $\mu_i\ge 0$, the leading behavior of the 2-nucleon
amplitude is $(QM)^L$.  
If one treats $M\simeq Q^0$, it
follows that perturbation theory is adequate for describing
the 2-nucleon system at low $Q$. 
In order to accommodate large scattering lengths and bound states near threshold,
as in the $\si$ and $\siii-\diii$ channels
one must conclude that $ M\sim 1/Q$ in this power counting
scheme.  
Thus the EFT calculation must be
an expansion in $\nu$, given by eqs.(\ref{wpo})(\ref{ourpo}).  
At leading order, $\nu=0$, one must sum up all ladder diagrams with
insertions of $V^{(2)}_0$ potentials with $\mu=0$, while  at
subleading order one includes one insertion of  $V^{(2)}_1$
and all powers of $V^{(2)}_0$, and so forth.
\begin{figure}[t]
\psfig{figure=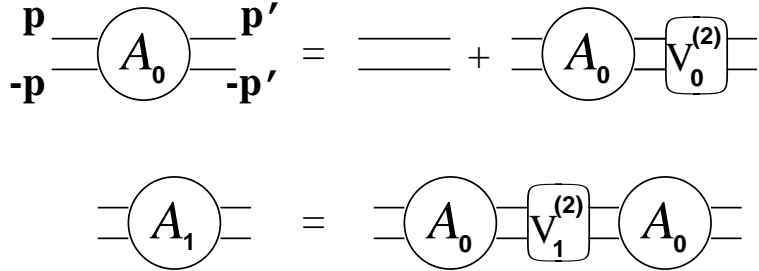,height=1.6in}
\caption{
The first two terms in the  EFT expansion of the Feynman amplitude
for nucleon-nucleon scattering in Weinberg's
power-counting scheme.  
The leading amplitude $\CA_0$
scales like $Q^0$ and consists of the sum of ladder diagrams with the leading
($\mu=0$) 2-nucleon potential $V^{(2)}_0$ at every rung.  
The subleading amplitude $\CA_1$ scales like $Q^1$. 
\label{wein_fig2}}
\end{figure}

\begin{figure}[t]
\psfig{figure=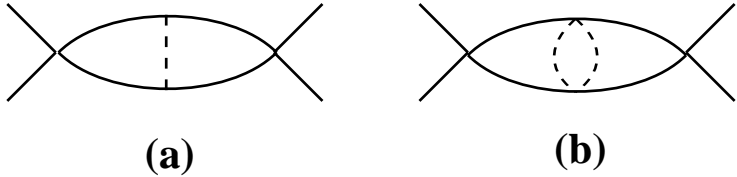,height=1.1in}
\caption{Graphs  with logarithmic divergences.
The divergence in graph (a) is proportional to $M^2 m_\pi^2$,
while graph (b) has a divergence  proportional to  $M^2 {\bf p}^4$. 
The solid lines are nucleons and the dashed lines are pions.
\label{wein_fig3}}
\end{figure}

At leading order in Weinbergs power-counting there are contributions to 
$V_0^{(2)}$ from both the local four-nucleon operators, $  C_{S,T}$ and 
from the exchange of a single potential pion,
giving a momentum space potential of 
\be
V_0^{(2)}({\bf p},{\bf p^\prime}) = C \ - \ \({g_A^2\over
2f_\pi^2}\){({\bf q} \cdot\sigma_1   {\bf q} \cdot\sigma_2)(\tau_1\cdot\tau_2)
\over ({\bf q}^2+m_\pi^2)\ }
\ \ \ \ ,
\label{eq:ope}
\ee
where $C$ denotes the  combination of $C_{S,T}$ appropriate for a
given spin-isospin channel.
The leading order amplitude results from summing the graphs 
shown in Fig.~(\ref{wein_fig2})
resulting  from this  potential, 
i.e. solving the Schrodinger equation.
In the $\si$ channel at two-loops in the ladder sum there is a  
logarithmic divergence in the graph shown in Fig.~(\ref{wein_fig3}a)
that  must be regulated.
In dimensional regularization the divergent part of this graph is 
\be
-{1\over \epsilon} {g_A^2 m_\pi^2 M^2\over 128\pi^2 f_\pi^2} C^2
\ \ \  ,
\label{eq:pole}
\ee
which requires a counterterm with a single insertion of the light 
quark mass matrix, for instance
\be
-{1\over 2} C_S^{(m)} 
Tr \left[m_q (\Sigma + \Sigma^\dagger)\right] 
\left(N^\dagger N\right)^2
-{1\over 2} C_T^{(m)} 
Tr \left[m_q (\Sigma + \Sigma^\dagger)\right] 
\left( N^\dagger \sigma N\right)^2
\ +\ ...
\ \ \ .
\ee
However, the coefficients of these operators must scale like $M^2$, and since
$ m_\pi^2 M^2 \sim Q^0$
these formally higher order operators in 
Weinberg's power-counting are required at leading order to absorb 
divergences in the time-ordered products of the leading order potential,
$V_0^{(2)}$.
Ignoring the multi-pion vertices arising from these operators, 
they can be re-absorbed into the leading operators with coefficients
$C_{S,T}$.
We are then in the  situation where there is no chiral expansion, multiple insertions
of the light quark mass matrix are not suppressed compared to leading order interactions,
but we do still have a momentum expansion in the $\si$ channel.
The graph shown in Fig.~(\ref{wein_fig3}b) requires counterterms involving  
$\nabla^4$ with  coefficients proportional to  $M^2$,
while this is not suppressed by $Q^4$
it is suppressed compared to the leading operators by $Q^2$.

\begin{figure}[t]
\psfig{figure=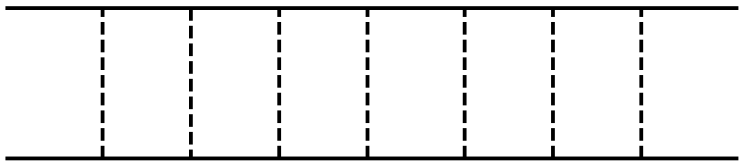,height=0.9in}
\caption{A contribution to the pion ladder sum, arising at leading 
order in Weinberg's power-counting.
The solid lines are nucleons and the dashed lines are pions.
\label{wein_fig4}}
\end{figure}

The situation is different in the $\siii-\diii$ channel and in higher partial waves.
A contribution to the leading order ladder sum is shown in Fig.~(\ref{wein_fig4}),
arising from seven potential pion exchanges, i.e a six-loop graph.
From our discussion it is straightforward to deduce that
this graph has a logarithmic divergence at order $(QM)^6$, and therefore, 
counterterms
involving $\nabla^6$ 
(i.e. operators up to and including those with orbital angular momentum $L=6$)
are required at leading order in the expansion.
Clearly, the same discussion can be made for an arbitrary number of 
potential pion exchanges, and therefore counterterms involving an arbitrary even 
number of $\nabla$'s are required.
This is a clear demonstration of the failure of Weinbergs power-counting.
Further, this conclusion is true for all regularization schemes 
and not just for dimensional regularization.
In using a momentum cut-off to regulate the theory, while truncating the expansion
of the potential at a given number of derivatives one will have terms that depend 
explicitly upon the cut-off $\Lambda$.
If the cut-off is kept small, then such terms do not have a large impact upon 
the ``goodness of fit'' to data and $\Lambda$ can be chosen to achieve the ``best fit''
\cite{KoMany,GPLa}.
The fact that the results depend upon $\Lambda$ demonstrates the presense of an incomplete
set of higher order terms.

As we will see in subsequent sections, the problem with Weinberg's power-counting
is having to identify $M\sim Q^{-1}$ and $C\sim Q^0$.
In fact consistent power-counting is obtained when we identify
$M\sim Q^0$ and $C\sim Q^{-1}$.

\section{A New Power Counting}
\label{sec:3}

Lets us begin by examining the general form of the amplitude
for nucleon scattering in a S-wave
\bea
\CA & = & {4\pi\over M} {1\over p \cot\delta - i p}
\ \ \ .
\label{eq:ampa}
\eea
From quantum mechanics it is well known that 
$p\cot\delta$ has a momentum
expansion for $p\ll\Lambda$ (the effective range expansion),
\be
p\cot\delta = -{1\over a} + {1\over 2}\Lambda^2\sum_{n=0}^\infty {r}_n
\({p^2\over \Lambda^2}\)^{n+1}
\ \ \ \ ,
\label{eq:erexp}
\ee
where $a$ is the scattering length, and 
$r_0$ is the effective range.
For scattering in the $\si$ and $\siii$ channels the scattering lengths
are found to be large, $a^{(\si)} = -23.714\pm 0.013\ {\rm fm}$ and 
$a^{(\siii)} = +5.425\pm 0.0014\ {\rm fm}$ respectively.
Expanding the expression for the amplitude in eq.(\ref{eq:ampa}) 
in powers of $p/\Lambda$ while retaining $ap$ to all orders gives
\be
\CA = -{4\pi\over M}{1\over (1/a + i p)}\[ 1 + {r_0/2 \over (1/a + ip)}p^2 +  
{(r_0/2)^2\over (1/a + ip)^2} p^4 + {(r_1/2\Lambda^2)\over (1/a + ip) } p^4 +\ldots\]
\label{eq:aexp2}
\ee
For $p>1/|a|$ the terms  in this expansion scale as 
$\{p^{-1}, p^0,p^1,\ldots\}$, and 
the expansion in the effective theory takes the form
\be
\CA=\sum_{n=-1}^\infty \CA_n 
\ \ \  ,\ \ \ \CA_n \sim p^n
\ \ \ \ .
\label{eq:ampbiga}
\ee

In the theory without pions we can explicitly compute the 
s-wave amplitude in each spin channel
to all orders in the momentum expansion,
\bea
\CA & = & -{  \sum C_{2n} p^{2n} 
\over
1 + M(\mu+ip)/4\pi \sum C_{2n} p^{2n} }
\ \ \ \  ,
\label{eq:answer}
\eea
where $C_{2n}$ is the coefficient of the $p^{2n}$ term in the lagrange density.
$\mu$ is the renormalization scale and we have used 
Power Divergence Subtraction (\pds)\cite{KSWb} to define the  theory.
A typical loop graph that appears in the amplitude has the form
\bea
I_n & \equiv & -i\left({\mu\over 2}\right)^{4-D} 
\int {{\rm d}^D q\over (2\pi)^D}
\, 
{\bf q}^{2n} 
\({i\over {E\over 2} + q_0 -{ {\bf q}^2\over 2M} + i\epsilon}\)
\({i\over {E\over 2} - q_0 -{ {\bf q}^2\over 2M} + i\epsilon}\)
\nonumber\\
& = & \left({\mu\over 2}\right)^{4-D}
\int {{{\rm d}}^{(D-1)}{\bf  q}\over (2\pi)^{(D-1)}}\, 
{\bf q}^{2n} \({1\over E  -{ {\bf q}^2\over M} + i\epsilon}\) 
\nonumber\\
& = & -M (ME)^n (-ME-i\epsilon)^{(D-3)/2 } \ \ \Gamma\({3-D\over 2}\)
{\left({\mu\over 2}\right)^{4-D} \over  (4\pi)^{(D-1)/2}}
\ \ \ \  ,
\label{eq:loopi}
\eea
where $D$ is the number of space-time dimensions.
In the \pds scheme the pole at $D=3$ is removed by adding a local counterterm
to the lagrange density, giving
\be
\delta I_n = -{M(ME)^n \mu\over 4\pi (D-3)}
\ \ \ ,
\ee
so that the sum of the loop graph and counterterm 
in $D=4$ dimensions is
\be
I_n^{PDS} = I_n + \delta I_n = - (ME)^n \left({M\over 4\pi}\right) (\mu + ip).
\label{eq:ipds}
\ee

\begin{figure}[t]
\psfig{figure=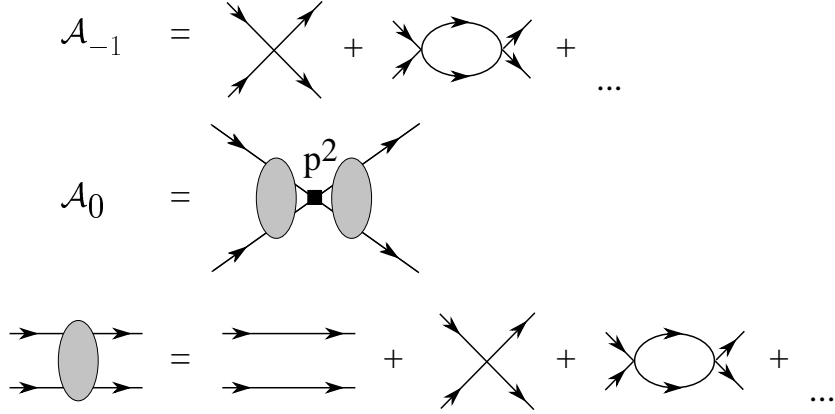,height=2.5in}
\caption{Leading and subleading contributions arising from local operators.
\label{KSW_fig2}}
\end{figure}

The amplitude $\CA$ is independent of the subtraction point $\mu$ and
this  determines the $\mu$ dependence of the coefficients, $C_{2n}$. 
In the \pds scheme one finds that for $\mu\gg 1/|a|$, 
the couplings $C_{2n}(\mu)$ scale as
\be
C_{2n}(\mu) \sim {4\pi \over M \Lambda^n \mu^{n+1}}\ ,
\label{eq:cscale}
\ee
so that if we take $\mu \sim p$, $C_{2n}(\mu)\sim 1/p^{n+1}$.  
A factor of $\nabla^{2n}$ at a vertex scales as $p^{2n}$, 
while each loop contributes a factor of  $p$.  
Therefore, the leading order contribution to the scattering amplitude $\CA_{-1}$  
scales as $p^{-1}$ and consists of the sum of bubble diagrams with $C_0$ vertices. 
Contributions  scaling as higher powers of $p$ come from perturbative 
insertions of derivative interactions, dressed to all orders by $C_0$.  
The first two  terms in the expansion
\bea
{\cal A}_{-1} & = &  { -C_0\over \left[1 + {C_0 M\over 4\pi} (\mu + ip)\right]}
\ \ \ ,\ \ \ 
{\cal A}_0  ={ -C_2 p^2\over \[1 + {C_0 M\over 4\pi}(\mu + ip)\]^2}
\ \ \ \ ,
\eea
correspond to the Feynman diagrams in Fig.~(\ref{KSW_fig2}).
A comparison with eq.(\ref{eq:aexp2}) gives
\bea
C_0(\mu) &=& {4\pi\over M }\({1\over -\mu+1/a}\)
\ \ \ , \ \ \ 
C_2(\mu) =  {4\pi\over M }\({1\over -\mu+1/a}\)^2 {{r}_0\over 2}
\ \ \ .
\label{eq:cvals}
\eea
The dependence of $C_{2n}(\mu)$ on
$\mu$ is determined by requiring  the amplitude be independent of the
renormalization scale  $\mu$.  
The physical parameters $a$, $ r_n$ enter as
boundary conditions on the resulting renormalization group (RG) 
equations. 
The beta function for each of the couplings $C_{2n}$ is defined by 
\be
\beta_{2n} \equiv \mu {{\rm d}C_{2n}\over {\rm d}\mu}
\ \ \ \ .
\label{eq:betadef}
\ee
In the \pds scheme, the $\mu$ dependence of the $C_{2n}$ coefficients enters
logarithmically or linearly, associated with simple $1/(D-4)$ or
$1/(D-3)$ poles respectively.  
\begin{figure}[t]
\psfig{figure=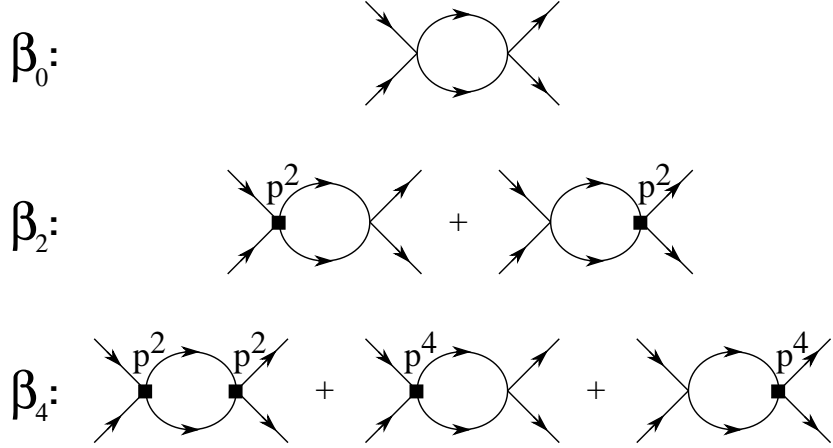,height=2.5in}
\caption{ Graphs contributing to the $\beta$-functions for $C_{2n}$
\label{KSW_fig3}}
\end{figure}
The functions $\beta_{2n}$  follow straightforwardly from
$\mu{d\over d\mu}(1/\CA)=0$ and for $C_0$ and  $C_2$ can be 
found from the one-loop
graphs in Fig.~(\ref{KSW_fig3})
\bea
\beta_0 & = &  {M\mu\over 4\pi} C_0^2
\ \ \ ,\ \ \ 
\beta_2 = 2 {M\mu\over 4\pi} C_0 C_2
\ \ \ .
\label{eq:beta024}
\eea
Integrating these equations relates the $C_{2n}$ coefficients at two different
renormalization scales $\mu$ and $\mu_0$. 
The solution for $C_0(\mu)$ and $C_2(\mu)$ 
with the boundary condition $C_0(0)=4\pi a/M$, 
is
\bea
C_0(\mu) & = &  {4\pi\over M }\({1\over -\mu+1/a}\)\ \ \ ,\ \ \ 
C_2(\mu) = C_2(\mu_0)\({C_0(\mu)\over C_0(\mu_0)}\)^2
\ \ \ \ ,
\label{eq:c2rg}
\eea
which when combined with the boundary condition,
$ C_2(0) = C_0(0) a r_0/ 2$, yields  $C_2(\mu)$ as given in eq.(\ref{eq:cvals}).
It is possible to solve the complete, coupled RG equations
for the leading small $\mu$ behavior of each of the coefficients $C_{2n}$ . 
The coefficients are 
\be
C_{2n}(\mu) = {4\pi \over M(-\mu+1/a)} \({r_0/2\over -\mu+1/a}\)^n + O(\mu^{-n})
\ \ \ ,
\ee
which has the scaling property in eq.(\ref{eq:cscale}).  
The leading behavior depends on the two parameters $a$ and $r_0$ 
encountered when solving for $C_0(\mu)$ and $C_2(\mu)$.  
This is due to the $C_{2n}$ couplings being driven primarily 
by lower dimensional interactions.

The inclusion of pions into the theory is straightforward.
While the coefficients of the local operators are renormalized,
and scale as powers of the renormalization scale $\mu$
(we use $Q \equiv \mu\sim p\sim m_q^{1/2}$),
the exchange of a single potential pion does not suffer from such
renormalizations and therefore pion exchange is
a sub-leading contribution, $Q^0$.
At the same order as the exchange of a potential pion is an
insertion of a $C_2$ operator and  a
single insertion of the quark mass matrix $m_q$. 
Ignoring isospin violation,  
these operators involving insertions of the 
light quark mass matrix with coefficients $D_2$
have the same structure as the $C_0$ operators.
For  $\mu\sim m_\pi$,
$C_0(\mu)\propto 1/\mu$, 
$C_2(\mu)\propto 1/\mu^2$ 
and $D_2(\mu)\propto  1/\mu^2$ 
for the $\si$ and $\siii$ channels.   
A  feature of the theory with pions is that this scaling 
behavior breaks down at low momentum, $p\sim 1/|a|$, 
and at sufficiently
high momentum.  
\begin{figure}[t]
\psfig{figure=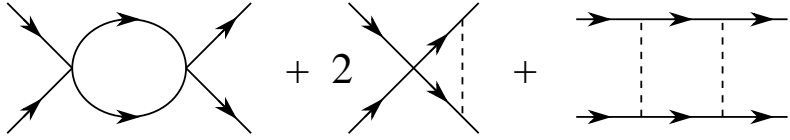,height=0.9in}
\caption{Contributions to the $\beta$-functions for $C_0$
in the theory with pions
\label{KSW_fig4}}
\end{figure}
The exact beta function for the $C_0$ coefficients from the graphs in
Fig.~(\ref{KSW_fig4}) are
\bea
\beta_0^{(\si)} & = & \mu {{\rm d}C_0^{(\si)}\over {\rm d}\mu\ \ } =  {M\mu\over 4\pi}
\left\{\(C_0^{(\si)}\)^2+ 2 C_0^{(\si)}{g_A^2\over 2 f^2}+ \({g_A^2\over 2
f^2}\)^2\right\}
\nonumber\\
\beta_0^{(\siii)}&  = & \mu {{\rm d}C_0^{(\siii)}\over {\rm d}\mu\ \ } =  {M\mu\over
4\pi} \left\{\(C_0^{(\siii)}\)^2+  2 C_0^{(\siii)}{g_A^2\over 2 f^2}+ 9 
\({g_A^2\over 2 f^2}\)^2\right\}
\ \ \  .
\label{eq:beta013}
\eea
Solving the RG equation in the $\si$ channel with the boundary
condition $C_0^{(\si)}(0)=4\pi a_1/M$, 
where $a_1$ is the $\si$ scattering length,  we find  for 
$\mu\gg 1/|a_1|$
\be
C_0^{(\si)}(\mu)\simeq -{4\pi\over M\mu}\(1+{\mu\over \Lambda_{NN}}\)\ ,
\ee
with
\bea
\Lambda_{NN} & = & {8\pi f^2\over g_A^2 M} \sim 300\ {\rm MeV}
\ \ \ ,
\eea
and therefore the power counting changes when $\mu \sim \Lambda_{NN}$.
The 
UV fixed point toward which $C_0^{(\si)}$ is driven largely cancels 
the $\delta$-function component
of the single potential pion exchange in the $\si$ channel.  
As a result, this power counting 
works only up  to $p\sim \Lambda_{NN}$
and the power counting in both channels is  expected 
to fail at momenta on the order of $\Lambda_{NN}$.  
We conclude from this discussion that the expansion parameter for this theory
is $\sim m_\pi/\Lambda_{NN}\sim {1\over 2}$, which is larger than one would like.

\section{NN scattering in the $\si$ Channel}
\label{sec:4}

\begin{figure}[t]
\psfig{figure=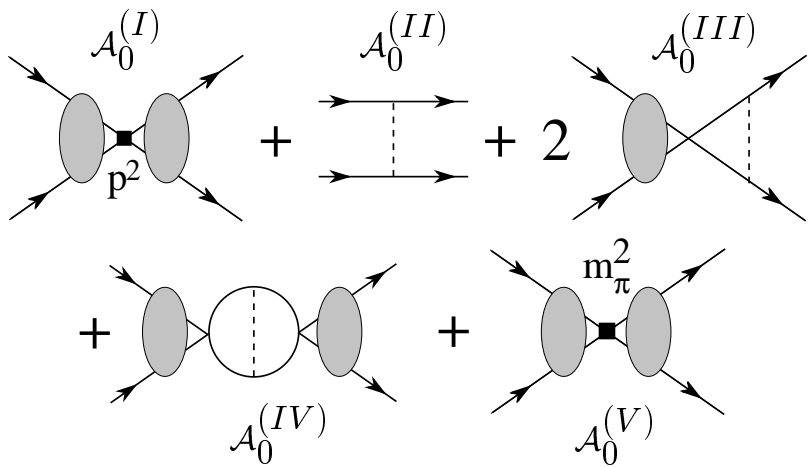,height=2.5in}
\caption{Graphs contributing to the subleading amplitude $\CA_{0}$.
The shaded ovals are defined in Fig.~(\ref{KSW_fig2}). 
\label{KSW_fig5}}
\end{figure}

Having established a consistent power-counting in the previous
sections we now apply it to $NN$ scattering in the $\si$ channel.
The amplitude at order $Q^{-1}$ and $Q^0$ determined 
from the graphs shown in
Fig.~(\ref{KSW_fig2}) and Fig.~(\ref{KSW_fig5})
are 
\bea
\CA_{-1} & = & 
-{ C^{(\si)}_0\over  1 + C^{(\si)}_0 {M\over 4\pi}  \left( \mu + i p \right) }
\ \ \ ,
\nonumber\\
 {\cal A}_0^{(I)} & = & 
-C_2^{(\si)} p^2
\left[ {\CA_{-1}\over C_0^{(\si)}  } \right]^2
\ \ \ ,
\nonumber\\
 {\cal A}_0^{(II)} &=&  \left({g_A^2\over 2f^2}\right) \left(-1 + {m_\pi^2\over
4p^2} \ln \left( 1 + {4p^2\over m_\pi^2}\right)\right)
\ \ \ \ ,
\nonumber\\
 {\cal A}_{0}^{(III)} &=& {g_A^2\over f^2} \left( {m_\pi M{\cal A}_{-1}\over 4\pi}
\right) \Bigg( - {(\mu + ip)\over m_\pi}
+ {m_\pi\over 2p} X(p,m_\pi)\Bigg)
\  \ \ \ ,
\nonumber\\
{\cal A}_0^{(IV)} &=& {g_A^2\over 2f^2} \left({m_\pi M{\cal A}_{-1}\over
4\pi}\right)^2 \Bigg(1 -\left({\mu + ip\over m_\pi}\right)^2
+ i X(p,m_\pi)  - \ln\left({m_\pi\over\mu}\right)  
\Bigg)
\ \ \ \ ,
\nonumber\\
{\cal A}_0^{(V)} &=& - D^{(\si)}_2 m_\pi^2 
\left[ {\CA_{-1}\over C_0^{(\si)}  }\right]^2
\ \ \ \  ,
\nonumber\\
X(p,m_\pi) & = & \tan^{-1} \left({2p\over m_\pi}\right) + {i\over 2} \ln
\left(1+ {4p^2\over m_\pi^2} \right)
\ \ \ \  .
\label{eq:amp1s0}
\eea
At order $Q^{-1}$ there is one unknown coefficient $C^{(\si)}_0$ that must be determined from 
data while at order $Q^0$ there are three unknown coefficients $C^{(\si)}_0, C^{(\si)}_2$ and
$D^{(\si)}_2$ that must be determined.
The graph giving $\CA_0^{(IV)}$ is divergent in four dimensions and therefore
gives rise to the logarithmic dependence on the renormalization scale $\mu$ 
in eq.~(\ref{eq:amp1s0})
(we have performed a finite subtraction, in addition to the usual 
\pds ).
In order for the expansion to converge, 
the leading term $\CA_{-1}$ must capture most of the 
scattering length, and hence the subtraction.
For this scattering channel we examine the phase shift $\delta$, 
and perform a perturbative expansion of $\delta$ in $Q$,
$\delta = \delta^{(0)} + \delta^{(1)} + \ldots $.
It is straightforward to relate the phase shift to the amplitude,
\bea
\delta & = & 
{1\over 2i} \ln\left( 1 + i {Mp\over 2\pi}\CA \right)
\ \ \ \ ,
\nonumber\\
\delta^{(0)} & = & 
{1\over 2i} \ln\left( 1 + i {Mp\over 2\pi}\CA_{-1} \right)
\ ,\qquad
\delta^{(1)} = {M p \over 4 \pi} \left( {\CA_0\over 1 + i {Mp\over 2\pi}\CA_{-1} }
\right)
\ \ \ \ .
\eea
\begin{figure}[t]
\psfig{figure=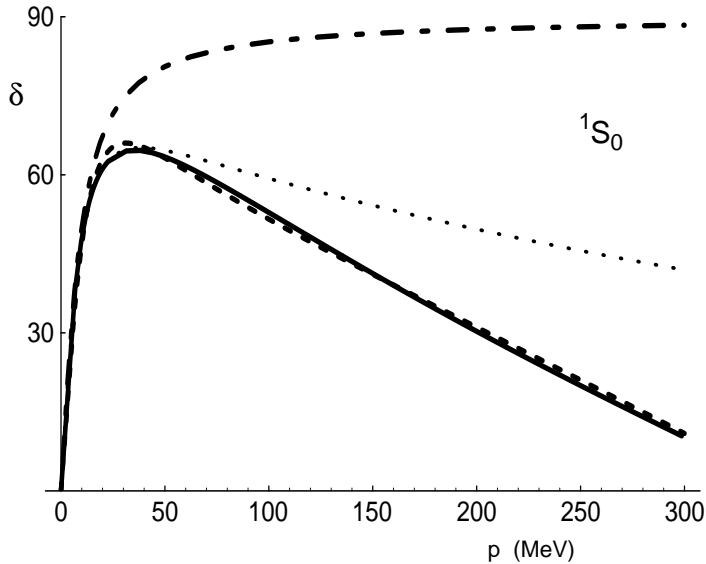,height=3.0in}
\caption{The phase shift $\delta$ for the $\si$ channel.
The dot-dashed curve is the  one parameter fit at order $Q^{-1}$,
that reproduces the scattering length.
The dashed curve corresponds to fitting $\delta$ 
between $0 < p < 200\ {\rm MeV}$, while the 
dotted curve corresponds to fitting the 
scattering length and effective range.
The solid line shows the results of the 
Nijmegen partial wave  analysis.
\label{KSW_fig6}}
\end{figure}

There are several ways to determine the coefficients from observables.
One fit we have performed is to the results of the 
Nijmegan partial-wave analysis \cite{nijmegen}
over a momentum range $p\le 200 \MeV$, from which we find for $\mu=m_\pi$
\be
C_0^{(\si)} =-3.34\fm^2
\ ,\qquad
D_2^{(\si)} =-0.42\fm^4
\ ,\qquad
C_2^{(\si)} =3.24\fm^4
\ \ \  ,
\label{eq:numfit}
\ee
giving the dashed curve plotted in  Fig.~(\ref{KSW_fig6}).
Fitting the scattering length and effective range gives 
(requiring $D_2(m_{\pi}) = 0$), for $\mu=m_\pi$
\be
C_0 =-3.63~{\rm fm}^2\ ,\qquad
C_2 =2.92~{\rm fm}^4
\ \ \ ,
\label{eq:dzfit}
\ee
and the dotted curve in  Fig.~(\ref{KSW_fig6}).
It is clear from Fig.~(\ref{KSW_fig6}) that the corrections to the 
leading order result become substantial above $\sim 200\  {\rm MeV}$
and we expect the expansion to become unreliable at momenta larger
than this value.
We chose to renormalize at $\mu=m_\pi$ for our numerical analysis, but
as  the amplitudes are explicitly $\mu$-independent we could have chosen 
any value of $\mu$, with $\Lambda_{NN}\gg\mu\gg 1/a$.  
The logarithm appearing in
the subleading amplitude suggests we choose $\mu\sim m_\pi$.

\section{NN scattering in the $\siii-\diii$ Channel}
\label{sec:5}

The analysis of scattering in the $\siii-\diii$ channel is a
straightforward extension of the analysis performed in the 
$\si$ channel.
The important difference is that the nucleons in the initial 
and final states with total angular momentum $J=1$ can be in an 
orbital angular momentum state of either $L=0$ or $L=2$.
The power counting for amplitudes that take the nucleons from a 
$\siii$-state to a $\siii$-state is identical to the analysis
in the $\si$-channel.
In fact, the expression for these amplitudes are exactly the same
with the appropriate substitution of coefficients.
Operators  between two $\diii$ states are not renormalized by the 
leading operators, which project out only $\siii$ states.
Further, they involve a total of four spatial derivatives, two on the incoming 
nucleons, and  two on the out-going nucleons.   
Therefore, such operators contribute
at order $Q^3$, and can be   neglected.
Consequently, amplitudes for scattering from an $\diii$ state into an $\diii$ state
are dominated by single potential pion exchange which contributes at order $Q^0$.
Operators connecting $\diii$ and $\siii$ states involve 2 spatial 
derivatives (acting on the $\diii$ state) 
and are renormalized by the leading  operators, but only on the 
$L=0$ ``side'' of the operator.
Therefore the coefficient of this operator, $C_2^{(\siii-\diii)}\sim 1/\mu$,
contributing at order $Q^1$ and  it can be neglected at order $Q^0$.
Thus,  mixing between $\diii$ and $\siii$ states is dominated by
single potential pion exchange dressed by a bubble chain of $C_0^{(\siii)}$ operators
and a parameter free prediction for this mixing 
exists at order $Q^0$.

We denote the amplitude at order $Q^n$ by $\CA_{n[LL^\prime]}$, 
where $L$ and $L^\prime$ are the 
initial and final orbital angular momenta.  
At leading order $Q^{-1}$ in the expansion there is a contribution only to the 
$\siii$ partial wave:
\begin{eqnarray}
\Czzm  = 
-{ C^{(\siii)}_0\over  1 + C^{(\siii)}_0 {M\over 4\pi}  \left( \mu + i p \right) 
}\ ,\ \ 
\Cztm\  = \  \Ctzm\  =\  \Cttm \ = \ 0
\  \ \ \ .
\label{eq:simib}
\end{eqnarray}
At order $Q^0$ there are contributions from  graphs of the same form as in 
the amplitude for $\si$ scattering, shown in Fig.~(\ref{KSW_fig5}).
Using the same identification of graphs  as in the $\si$ channel, 
${\cal A}_{0[L,L^{\prime}]}={\cal A}_{0[L,L^{\prime}]}^{(I)} +...$,
we find that 
\bea
\Czzz^{(I)}  & = & 
-C_2^{(\siii)} p^2
\left[ {\Czzm\over C_0^{(\siii)}  } \right]^2
\ ,\qquad
\Cztz^{(I)} \  = \  \Ctzz^{(I)} \ =\  \Cttz^{(I)} 
\ =\ 0
\ ,
\nonumber\\
\Czzz^{(II)} \ & = &\  
-{g_A^2\over 2 f^2} \left[ 1 - {m_\pi^2\over 2 p^2} Q_0(z)\right]
\ \ \ ,
\nonumber\\
\Cztz^{(II)} \ & = &\  \Ctzz^{(II)} \ =\ 
-{g_A^2  \over \sqrt{2}  f^2} 
\left[ Q_0(z) + Q_2(z)-2 Q_1(z) \right]
\ \ \ ,
\nonumber\\
\Cttz^{(II)} \ & = &\  
-{g_A^2  m_\pi^2\over 4 f^2 p^2} 
\left[ Q_2(z) + {6 p^2\over 5m_\pi^2}\left(Q_1(z)-Q_3(z)\right)\right]
\ ,
\nonumber\\
\Czzz^{(III)} &=& {g_A^2\over f^2} \left( {m_\pi M \Czzm\over 4\pi}
\right)\ 
\Bigg( - {(\mu + ip)\over m_\pi}
+ {m_\pi\over 2p} X(p,m_\pi)\Bigg)
\ \ \ ,
\nonumber\\
\Cztz^{(III)} \ & = & \ \Ctzz^{(III)} 
\nonumber\\
& = & 
{g_A^2\over \sqrt{2} f^2}
\left({ M  \Czzm\over 4 \pi}\right) p^2 
\left[ -{3 m_\pi^3\over 4 p^4} 
+ {m_\pi^2\over 8 p^5} (3 m_\pi^2 + 4 p^2)
\tan^{-1}\left({2 p\over m_\pi}\right)
\right.
\nonumber\\
& & \left.
\qquad\qquad + i \left( -{3 m_\pi^2\over 4 p^3} 
+ {1\over 2 p} 
+ {m_\pi^2\over 4 p^3} 
\left( 1+ {3 m_\pi^2\over 4 p^2} \right)
\log\left(1+{4 p^2\over m_\pi^2}\right)
\right)
\right]
\ \ \ ,
\nonumber\\
\Cttz^{(III)} \ & =& \  0
\ \ ,
\nonumber\\
\Czzz^{(IV)} &=& 
{g_A^2\over 2f^2} 
\left({m_\pi M \Czzm\over 4\pi}\right)^2 
 \Bigg(1 -\left({\mu + ip\over m_\pi}\right)^2
+ i X(p,m_\pi)  - \ln\left({m_\pi\over\mu}\right)  \Bigg)\ ,
\nonumber\\
\Cztz^{(IV)} \ &=& \ \Ctzz^{(IV)} \ =\ 
\Cttz^{(IV)} \ = \  0
\   ,
\nonumber\\
\Czzz^{(V)} & =&  - D^{(\siii)}_2 m_\pi^2 
\left[ {\Czzm\over C_0^{(\siii)}  }\right]^2\ ,\qquad
\Cztz^{(V)} \ = \ \Ctzz^{(V)} \ =\ 
\Cttz^{(V)} \ = \  0
\  \ \ \  ,
\eea
where $z = 1+m_\pi^2/(2 p^2)$ and  $Q_k (z)$ denotes
the $k$-th order irregular Legendre function.
Again part of the subtraction point independent contribution to $\Czzz^{(IV)}$
has been absorbed into $\Czzz^{(V)}$.
The S-matrix
in this channel is expressed in terms of two phase shifts,
$\delta_{0}$ and 
$\delta_{2}$, and a mixing angle $\varepsilon_1$,
\bea
S =  1\ +\ i{pM \over 2\pi}{\cal A}=\left( \begin{array}{ll}
e^{2i\delta_{0}} \cos 2\varepsilon_{1}  & ie^{i(\delta_{0} + \delta_{2})}
\sin 2\varepsilon_1\\
ie^{i(\delta_{0} + \delta_{2})} \sin 2\varepsilon_1 & e^{2i\delta_{2}} \cos
2\varepsilon_1 \end{array} \right) 
\  \ \ \  ,
\eea
and like the $\si$ channel we will expand $S$ order by order
in $Q$.

As the $\siii-\diii$ mixing parameter
has vanishing contribution at order $Q^{0}$ it  
starts at order $Q^1$, the same holds true for $\delta_2$.
Writing each of the parameters as an expansion in $Q$,
\bea
\delta_0 & = &  \delta_0^{(0)}+\delta_0^{(1)}+ ...
\ ,\ 
\delta_2 =  \delta_2^{(0)}+\delta_2^{(1)}+ ...
  \ ,\  
\varepsilon_{1} =  \varepsilon_{1}^{(0)}+ \varepsilon_{1}^{(1)}+ ...
\ ,
\eea
it follows that 
\bea
\delta_0^{(0)}  =  -{i\over 2} 
\log\left[ 1 + i {p M\over 2\pi} \ \Czzm\right]
\ ,\qquad 
\delta_0^{(1)} = {p M\over 4\pi} 
{ \Czzz\over   1 + i {p M\over 2\pi} \ \Czzm }\ ,
\eea
\bea
\varepsilon_{1}^{(0)}  =  0
\ ,\qquad
\varepsilon_{1}^{(1)} = {p M\over 4\pi} { \Cztz   \over
\sqrt{1 + i {p M\over 2\pi} \ \Czzm } }\ ,
\eea
\bea
\delta_{2}^{(0)} =  0
\ ,\qquad
\delta_{2}^{(1)} = {p M\over 4\pi} \ \Cttz 
\  .
\eea
\begin{figure}[t]
\psfig{figure=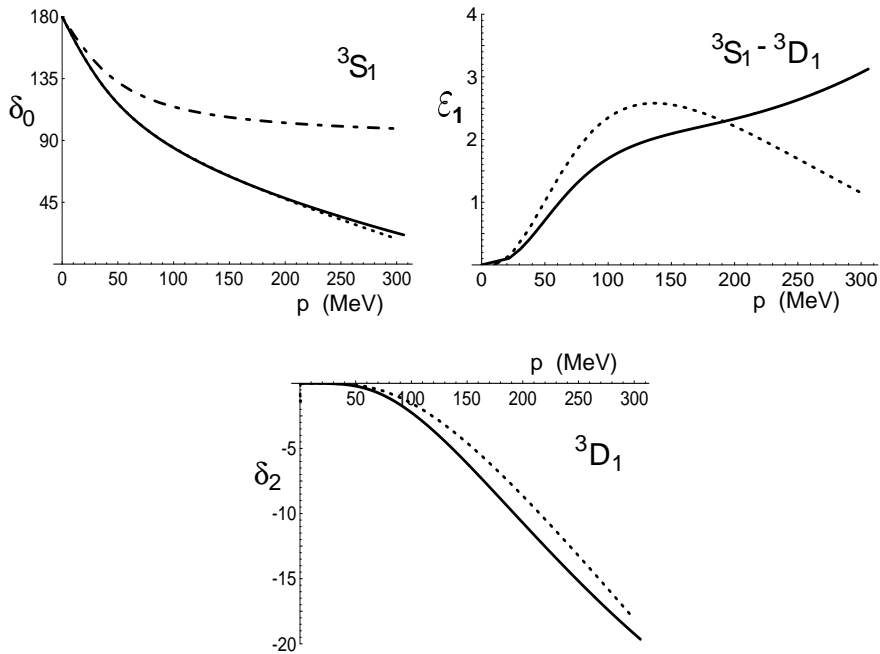,height=3.5in}
\caption{The phase shifts $\delta_0$, $\delta_2$ and  mixing parameter
$\varepsilon_1$ for the $\siii-\diii$ channel. 
The solid line denotes the results of the  Nijmegen partial wave  analysis.
The dot-dashed curve is the fit at order $Q^{-1}$ for $\delta_0$, while
$\delta_2 = \varepsilon_1 = 0$ at this order.
The dashed curves are the results of the order $Q^{0}$  fit of $\delta_0$ 
to the partial wave analysis over the momentum range $p\le 200\ \MeV$.
\label{KSW_fig7}}
\end{figure}
Fitting the parameters $C_0^{(\siii)}$,  $C_2^{(\siii)}$
and  $D_2^{(\siii)}$  to the phase shift $\delta_0$
over the momentum range $p\le 200 \MeV$ yields, at $\mu=m_\pi$
\bea
C_0^{(\siii)} =-5.51\fm^2\ ,\ 
D_2^{(\siii)} =1.32\fm^4\ ,\ 
C_2^{(\siii)} =9.91\fm^4
\  .
\label{eq:numfitc}
\eea
The dashed curves in Fig.~(\ref{KSW_fig7})  show the  phase shifts $\delta_0$, $\delta_2$
and mixing parameter $\varepsilon_1 $
compared to the  Nijmegen partial wave  analysis \cite{nijmegen} for this set of
coefficients.
There are no free parameters at this order in either $\varepsilon_1 $ or $\delta_2 $
once $C_0^{(\siii)}$ has been determined from $\delta_0$.

\section{Higher Order Effects}
\label{sec:6}

A local operator that connects an orbital angular momentum $L$ state
with an  orbital angular momentum $L^\prime$ state involves at least
$L+L^\prime$ spatial derivatives.
If either $L$ or $L^\prime$ but not both correspond to an $S$-wave 
then the operator enters at order $Q^{L+L^\prime-1}$.
However, if neither $L$ nor $L^\prime$ correspond to an $S$-wave then 
the operator contributes at order $Q^{L+L^\prime}$ when $L,L^\prime$ is
odd and at order $Q^{L+L^\prime-1}$ when $L,L^\prime$ is even. 
The contribution of pions is at order $Q^0$, and is therefore
the leading contribution to all non $S$-wave to $S$-wave 
scattering amplitudes.

Implicit in all previous discussions   we have  
been retaining only the contribution to diagrams from the nucleon poles.
While these are the leading contribution 
by factors of the nucleon mass, there are higher order contributions arising
from other poles, when present.
The leading contribution from the pion radiation regime 
(hence the term radiation pions, in analogy with the terminology in 
non-relativistic gauge 
theories\cite{LukeManohar2}$^-$\cite{HGr})
does not contribute to the $\beta$-functions of the $C_0(\mu)$ due to the  
explicit factors of $m_\pi$ but to the $\beta$-functions of the $D_2(\mu)$.
Further, such contributions give rise to mixing between the operators 
in the $\si$ and $\siii$ channels,
\bea
\beta_{D_2^{(\si)}}^{(rad)} & = & +\ {3 g_A^2  
\over 4 \pi^2 f^2} \left( C_0^{(\siii)} - C_0^{(\si)} \right)
\ \ \ ,
\nonumber\\
\beta_{D_2^{(\siii)}}^{(rad)} & = & -\ {3 g_A^2  
\over 4 \pi^2 f^2} \left( C_0^{(\siii)} - C_0^{(\si)} \right)
\ .
\eea

A comment on the role of   baryonic resonances is appropriate.
The impact of the $\Delta$ resonance has been determined with Weinberg's 
power-counting
in two different prescriptions\cite{KoMany,Sa96}
\footnote{The $\Delta (1232)$ and other  baryon resonances have been 
consistently included in the single nucleon sector\cite{JMhung}.}.
It is found not to play an important role in $NN$ scattering as the  
mass scale that sets the size of its contribution is 
$\sqrt{M (M_\Delta-M)}\sim 500 \ {\rm MeV}$.
This scale is higher than the scale at which the theory breaks down, 
$\Lambda_{NN}$, and so 
it is appropriate not to include the baryonic resonances, until the theory above
the scale $\Lambda_{NN}$ is constructed.

\section{Conclusions}
\label{sec:7}

After several years of investigation we now understand the 
limitations of Weinberg's  power-counting\cite{Weinberg1}.
In theories with large scattering lengths  as arise in
$NN$ scattering in the  $\si$ and $\siii-\diii$ channels,
simply counting derivatives or number of insertions of the light 
quark mass matrix does not correspond directly to the degree
of suppression of an operator.
A new power-counting scheme has been introduced\cite{KSWb} 
which allows simple identification of graphs that contribute 
at a given order.
In the new power-counting scheme  pions are sub-leading 
compared to the leading local four-nucleon operators
and  can be treated in perturbation theory.
The regulation problems found in Weinberg's theory do not arise.
Explicit computation of $NN$ scattering in the 
$\si$ and $\siii-\diii$ channels to sub-leading order has been 
presented.
Most impressive perhaps is the parameter-free prediction
of the $\siii-\diii$ mixing parameter $\varepsilon_1$
which agrees well with the Nijmegen phase shift analysis.

The future looks extremely promising for a systematic 
effective field theory analysis
of nuclear physics.
The short term program will be to examine  the two-body systems in detail,
i.e. the properties of the deuteron\cite{KSWnew} 
and inelastic processes.
In the long-term one hopes to make progress in many-body systems.
Some impressive results have already been obtained\cite{Bvk}
in the three-body systems.

\bigskip\bigskip

I would like to thank my co-organizers of this meeting,
Ryoichi Seki and Bira van Kolck,
who did all the hard work and who convinced El Nino to stay away.
I would also like to thank my collaborators, David Kaplan and Mark Wise.
This work is supported in part by 
Department of Energy Grant DE-FG03-97ER41014.



\section*{References}
\begin{thebibliography}{99}
\bibitem{Weinberg1}S. Weinberg,
\Journal{\PLB}{251}{288}{1990};
\Journal{\NPB}{363}{3}{1991};
\Journal{\PLB}{295}{114}{1992}.


\bibitem{KoMany} C. Ordonez and U. van Kolck, 
\Journal{\PLB}{291}{459}{1992};
C. Ordonez, L. Ray and  U. van Kolck, 
\Journal{\PRL}{72}{1982}{1994} ;
\Journal{\PRC}{53}{2086}{1996} ;  
U. van Kolck, 
\Journal{\PRC}{49}{2932}{1994} .

\bibitem{Parka} T.S.  Park, D.P.  Min and M. Rho,
\Journal{\PRL}{74}{4153}{1995} ;
\Journal{\NPA}{596}{515}{1996}.

\bibitem{KSWa} D.B. Kaplan, M.J. Savage and M.B. Wise,
\Journal{\NPB}{478}{629}{1996}, 
{\tt nucl-th/9605002}.


\bibitem{CoKoM} T. Cohen, J.L. Friar, G.A. Miller and 
U. van Kolck, 
\Journal{\PRC}{53}{2661}{1996}.

\bibitem{DBK} D. B. Kaplan, 
\Journal{\NPB}{494}{471}{1997}.

\bibitem{cohena}T.D. Cohen, 
\Journal{\PRC}{55}{67}{1997}.
D.R. Phillips and T.D. Cohen, 
\Journal{\PLB}{390}{7}{1997}.  
K.A. Scaldeferri, D.R. Phillips, C.W. Kao and T.D. Cohen,
\Journal{\PRC}{56}{679}{1997}.
S.R. Beane, T.D. Cohen and D.R. Phillips,
nucl-th/9709062.
 

\bibitem{Fria} J.L. Friar,  Few Body Syst. {\bf 99}, 1 (1996),
{\tt nucl-th/9607020}. 

\bibitem{Sa96} M.J. Savage, 
\Journal{\PRC}{55}{2185}{1997},
{\tt nucl-th/9611022}. 

\bibitem{LMa} M. Luke and A.V. Manohar, 
\Journal{\PRD}{55}{4129}{1997},  
{\tt hep-ph/9610534 }.

\bibitem{GPLa} G.P. Lepage, {\tt nucl-th/9706029},
Lectures given at 9th Jorge Andre Swieca Summer School: 
Particles and Fields, Sao Paulo,
Brazil, 16-28 Feb 1997.

\bibitem{Adhik} S.K. Adhikari and A. Ghosh, 
J. Phys. {\bf A30}, 6553 (1997).

\bibitem{RBMa}  K.G. Richardson, M.C. Birse and J.A. McGovern,
{\tt hep-ph/9708435}.
 
\bibitem{Bvk} P.F. Bedaque and U. van Kolck, 
{\tt nucl-th/9710073};
P.F. Bedaque, H.-W. Hammer and U. van Kolck,
{\tt nucl-th/9802057}.

\bibitem{aleph}
U. van Kolck, Talk given at
Workshop on Chiral Dynamics: Theory and Experiment (ChPT 97), Mainz,
Germany, 1-5 Sep 1997. 
{\tt hep-ph/9711222 }

\bibitem{Parkb} T.S.  Park, K. Kubodera,  D.P.  Min and M. Rho, 
{\tt hep-ph/9711463}.

\bibitem{KSWb} D.B. Kaplan, M.J. Savage and M.B. Wise, 
{\tt nucl-th/9801034}, {\it to appear in Phys. Lett. B};
{\tt nucl-th/9802075}, {\it submitted to Nucl. Phys. B}.

\bibitem{Gegelia} J. Gegelia,
{\tt  nucl-th/9802038}.

\bibitem{steelea} J.V. Steele and R.J. Furnstahl, 
{\tt nucl-th/9802069}.

\bibitem{GeorgiManohar}    A. Manohar and  H. Georgi,
\Journal{\NPB}{234}{189}{1984}.

\bibitem{nijmegen} 
V.G.J. Stoks, R.A.M. Klomp, C.P.F. Terheggen and J.J. de Swart,
\Journal{\PRC}{49}{2950}{1994},
{\tt nucl-th/9406039}.

\bibitem{LukeManohar2}     M. Luke and A.V. Manohar,
\Journal{\PLB}{286}{348}{1992},
{\tt hep-ph/9205228}.

\bibitem{Labelle}  P. Labelle, 
{\tt hep-ph/9608491}.

\bibitem{GrinsteinRothstein} B. Grinstein and I.Z. Rothstein, 
\Journal{\PRD}{57}{78}{1998}.
{\tt hep-ph/9703298}.

\bibitem{LukeSavage} M. Luke and M.J. Savage,
\Journal{\PRD}{57}{413}{1998},
{\tt hep-ph/9707313}.

\bibitem{HGr}H. Griesshammer, 
{\tt hep-ph/9712467}.

\bibitem{JMhung}E. Jenkins and A.V. Manohar,
talk presented at the Workshop on Effective Field Theories of
the Standard Model, Dobogoko, Hungary, Aug 1991.
Published in
{\it Effective Field Theories of the Standard Model},
ed. U. Meissner, World Scientific, Singapore (1992).

\bibitem{KSWnew} D.B. Kaplan, M.J. Savage and M.B. Wise,
nucl-th/9804032.

\end{thebibliography}
\end{document}